\documentclass{article}
\usepackage{arxiv}

\usepackage[T1]{fontenc}
\usepackage[utf8]{inputenc}
\usepackage{newtxtext,newtxmath}
\usepackage{amsmath}
\usepackage{booktabs}
\usepackage{microtype}
\usepackage{graphicx}
\usepackage{pgfplots}
\pgfplotsset{compat=1.18}
\usepackage[numbers,sort&compress]{natbib}
\usepackage{xcolor}
\usepackage{hyperref}
\usepackage{url}

\hypersetup{
  colorlinks=true,
  linkcolor=blue,
  citecolor=blue,
  urlcolor=blue,
  pdftitle={End-to-End Neural Shrinkage of Indefinite Pairwise Correlation Matrices for Small-Cap-Inclusive Portfolios},
  pdfauthor={Christian Bongiorno and Lorenzo Villassero},
  pdfsubject={Preprint},
  pdfkeywords={portfolio optimization, covariance estimation, missing data, nonlinear shrinkage, random matrix theory, market impact, neural networks}
}

\providecommand{\Description}[1]{}
\renewcommand{\headeright}{Preprint}
\renewcommand{\undertitle}{Preprint}
\renewcommand{\shorttitle}{Neural Shrinkage of Indefinite Pairwise Correlations}

\title{End-to-End Neural Shrinkage of Indefinite Pairwise Correlation Matrices for Small-Cap-Inclusive Portfolios}

\author{
  Christian Bongiorno \\
  \normalfont Universit\'e Paris-Saclay, CentraleSup\'elec \\
  \normalfont Math\'ematiques et Informatique pour la Complexit\'e et les Syst\`emes \\
  \normalfont 91190 Gif-sur-Yvette, France \\
  \normalfont \texttt{christian.bongiorno@centralesupelec.fr}
  \And
  Lorenzo Villassero \\
  \normalfont Universit\'e Paris-Saclay, CentraleSup\'elec \\
  \normalfont Math\'ematiques et Informatique pour la Complexit\'e et les Syst\`emes \\
  \normalfont 91190 Gif-sur-Yvette, France
}

\date{August 2026}

\begin{document}
\maketitle

\begin{abstract}
Small-cap-inclusive equity universes contain recently listed and intermittently traded securities, so enforcing a common look-back discards a substantial fraction of the available information. Pairwise-complete estimation preserves the longest overlap for each asset pair, but the resulting correlation matrix can be indefinite because its entries are computed on different samples. This prevents direct use in Markowitz optimization and falls outside the assumptions of standard random-matrix shrinkage. We adapt a rotation-invariant neural covariance estimator to this setting. The model computes mask-aware marginal moments and a pairwise correlation matrix proxy, processes its signed spectrum, and uses a bidirectional gated recurrent unit conditioned on factor-aligned effective sample lengths derived from the overlap matrix and eigenvector loadings. It maps all eigenvalues, including negative ones, to a positive inverse spectrum. The reconstructed covariance is positive definite and is trained end-to-end to minimize five-day realized global-minimum-variance risk. We evaluate 26 expanding-window models from 2000 to 2025 on up to 1,500 U.S. equities in a closing-auction simulator with point-in-time selection, commissions, financing, corporate actions, and market impact. Across the 26-year out-of-sample period, the neural estimator reduces annualized five-day volatility by approximately 20\% and increases the Sharpe ratio by approximately 40\% relative to the next-best covariance estimator. These improvements are consistent across realized risk, risk-adjusted performance, and drawdown control, remain after the modeled execution frictions, and are supported by a 99.9\% Model Confidence Set that retains only the neural estimator.
\end{abstract}


\section{Introduction}

We consider Markowitz portfolio construction when the investable universe is determined, for example, by an investment mandate or a separate alpha signal. This approach requires estimating a covariance matrix from a rectangular return panel in which all selected assets are observed over a common look-back window. This condition is often violated when the universe includes recent initial public offerings, lower-capitalization securities with intermittent trading, temporarily suspended stocks, or assets with incomplete historical coverage. In these settings, excluding a security solely because its return history is shorter or fragmented would alter the prescribed investment universe and is not a negotiable solution. 

A natural alternative is to estimate each covariance or correlation entry from the maximum history shared by the corresponding pair. This pairwise-complete construction uses more information than complete-case deletion and is consistent with the economic motivation for studying unequal return histories \cite{stambaugh1997histories}. Its numerical consequence is less convenient: because different entries are estimated from different subsets of dates, the assembled matrix is not necessarily positive semidefinite. Negative eigenvalues imply that the matrix can assign negative variance to some portfolios and therefore cannot be interpreted as a valid covariance matrix. Several methods seek to address this problem. Nearest-correlation approaches project the indefinite estimate onto the positive-semidefinite cone \cite{higham2002nearest}, whereas regression and factor-imputation methods construct a valid covariance matrix directly from the incomplete panel \cite{gramacy2008covariance,cahan2023factor}. These methods, however, do not jointly address the second challenge faced by large universes. Let $n$ denote the number of assets and $\Delta t_\mathrm{in}$  the length of a common estimation window. When the concentration ratio $q:=n/\Delta t_\mathrm{in}$  is not negligible, empirical eigenvalues are strongly affected by sampling noise and require substantial shrinkage \cite{bun2017cleaning}.

Existing Random Matrix Theory (RMT) based shrinkage methods are derived for covariance matrices computed from a common rectangular return panel. In this setting, all entries share the same sample size and the empirical covariance matrix is positive semidefinite. Pairwise-complete estimation violates both conditions. Each entry is computed from a different number of overlapping observations, and the assembled matrix may contain negative eigenvalues. Standard RMT results therefore do not characterize the spectral statistics of this estimator and do not provide a principled rule that simultaneously repairs the negative eigenvalues and shrinks the positive spectrum. To our knowledge, these two operations have not been combined within a consistent analytical framework.

Bongiorno et al. \cite{bongiorno2025rienet} showed that a dimension-agnostic recurrent network can learn a rotation-invariant eigenvalue map directly from a realized Global Minimum Variance (GMV) objective. This approach does not require an explicit analytical model for the empirical spectrum. However, it cannot be applied unchanged when the input matrix may be indefinite and its entries are based on heterogeneous overlap lengths. We extend it to indefinite pairwise cross-moment matrices by providing the network with the signed eigenvalues and factor-specific measures of the available sample information. 

Specifically, we condition the spectral map on factor-aligned effective sample lengths derived from pairwise overlaps and squared eigenvector loadings. The network transforms the signed spectrum into a positive inverse spectrum and is trained end-to-end to minimize future GMV variance. For evaluation, the resulting covariance estimates are passed to the same long-only GMV optimizer used for all benchmarks.

The empirical analysis evaluates whether the statistical gains survive realistic implementation costs in a universe extended toward lower-capitalization equities. We use a point-in-time universe of up to 1,500 U.S. stocks, retain short and fragmented return histories, and rebalance every five sessions from 2000 through 2025. Each strategy is evaluated in a continuous broker simulation that accounts for integer-share holdings, closing-auction execution, commissions, regulatory fees, dividends, splits, corporate-action settlements, and financing. Market impact follows a square-root closing-auction specification with coefficients that vary across market-capitalization groups and listing exchanges \cite{goyal2026impact}. This execution layer tests whether reductions in estimated portfolio risk remain economically relevant once concentration, turnover, and trade size are translated into realized trading costs.


\section{Indefinite Pairwise Risk Estimation}

Let $\mathbf{R}\in\mathbb{R}^{\Delta t_{\mathrm{in}}\times n}$ denote a return panel and $\mathbf{M}\in\{0,1\}^{\Delta t_{\mathrm{in}}\times n}$ its validity mask, where $M_{ti}=1$ indicates that $R_{ti}$ is observed. The overlap matrix is
\begin{equation}
\mathbf{T}=\mathbf{M}^{\top}\mathbf{M},
\qquad
T_{ij}=\sum_{t=1}^{\Delta t_{\mathrm{in}}}M_{ti}M_{tj}.
\label{eq:overlap}
\end{equation}
The marginal sample moments are
\begin{equation}
\hat{\mu}_i
=
\frac{1}{T_{ii}}
\sum_{t=1}^{\Delta t_{\mathrm{in}}}M_{ti}R_{ti},
\qquad
\hat{\sigma}_{i}^{2}
=
\frac{1}{T_{ii}}
\sum_{t=1}^{\Delta t_{\mathrm{in}}}
M_{ti}\bigl(R_{ti}-\hat{\mu}_i\bigr)^{2}.
\label{eq:marginal_mle}
\end{equation}
Define the marginally standardized returns
\begin{equation}
Z_{ti}
=
M_{ti}
\frac{R_{ti}-\hat{\mu}_i}{\hat{\sigma}_i}.
\label{eq:marginal_standardization}
\end{equation}
For $T_{ij}>0$, the marginally standardized pairwise cross-moment matrix entry is
\begin{equation}
C_{\cap,ij}
=
\frac{1}{T_{ij}}
\sum_{t=1}^{\Delta t_{\mathrm{in}}}Z_{ti}Z_{tj},
\label{eq:pairwise_correlation}
\end{equation}
with entries having no overlap set to zero. Unlike overlap-specific Pearson centering, Eqs.~\eqref{eq:marginal_mle}--\eqref{eq:pairwise_correlation} estimate each asset's marginal moments once and reuse them for every pair. With
\begin{equation}
\hat{\mathbf{D}}
=
\operatorname{Diag}
\bigl(\hat{\sigma}_1,\ldots,\hat{\sigma}_n\bigr),
\qquad
\boldsymbol{\Sigma}_{\cap}
=
\hat{\mathbf{D}}\mathbf{C}_{\cap}\hat{\mathbf{D}},
\label{eq:pairwise_covariance}
\end{equation}
$\boldsymbol{\Sigma}_{\cap}$ is the marginally centered pairwise cross-moment matrix. 

Although $\mathbf{C}_{\cap}$ is symmetric with unit diagonal, its entries use different subsets and different denominators. It is therefore not necessarily positive semidefinite and may have negative eigenvalues \cite{higham2002nearest}. We write
\begin{equation}
\begin{aligned}
\mathbf{C}_{\cap}
&=
\mathbf{Q}_\cap\boldsymbol{\Lambda}_{\cap}\mathbf{Q}_\cap^{\top},\\
\boldsymbol{\Lambda}_{\cap}
&=
\operatorname{Diag}
\bigl(\lambda_{\cap,1},\ldots,\lambda_{\cap,n}\bigr),
\qquad
\lambda_{\cap,1}\leq\cdots\leq\lambda_{\cap,n}.
\end{aligned}
\label{eq:pairwise_eigendecomposition}
\end{equation}
Since $\hat{\mathbf{D}}$ is nonsingular, $\boldsymbol{\Sigma}_{\cap}$ and $\mathbf{C}_{\cap}$  are congruent and therefore have the same inertia by Sylvester's law of inertia \cite[p.~282]{horn2013matrix}. Marginal scaling alone therefore cannot repair an indefinite pairwise estimate.

The conventional pairwise-complete estimator recomputes means and variances on each overlap. Its correlations are therefore bounded in $[-1,1]$, although the assembled matrix may still be indefinite \cite{higham2002nearest}. By contrast, Eqs.~\eqref{eq:marginal_mle}--\eqref{eq:pairwise_covariance} use marginal moments estimated from each asset's full available history, so that $\mathbf{C}_{\cap}$ and $\boldsymbol{\Sigma}_{\cap}$ collect marginally standardized pairwise cross-moments. Provided that marginal and overlap-conditioned moments converge to the same limits, the two constructions are asymptotically equivalent. We adopt marginal standardization to retain the available information for forecasting \cite{stambaugh1997histories} and to remain consistent with the standard RMT convention of applying a common normalization to the data matrix \cite{bai2010spectral,bun2017cleaning}. A valid correlation matrix is subsequently reconstructed by our neural network. For brevity, we refer below to $\mathbf{C}_{\cap}$ and $\boldsymbol{\Sigma}_{\cap}$ as the pairwise correlation and covariance estimates, respectively.

Once regularization yields a positive-definite covariance estimate $\widehat{\boldsymbol{\Sigma}}$, the unconstrained GMV portfolio is
\begin{equation}
\mathbf{w}
=
\frac{
\widehat{\boldsymbol{\Sigma}}^{-1}\mathbf{1}
}{
\mathbf{1}^{\top}
\widehat{\boldsymbol{\Sigma}}^{-1}\mathbf{1}
}.
\label{eq:gmv}
\end{equation}
The proposed method sets $\widehat{\boldsymbol{\Sigma}}=\boldsymbol{\Sigma}_{\mathrm{NN}}$, whereas $\boldsymbol{\Sigma}_{\cap}$ is its indefinite pairwise precursor. At deployment, each valid estimate is used in the common long-only problem
\begin{equation}
\min_{\mathbf{w}}\ 
\mathbf{w}^{\top}\widehat{\boldsymbol{\Sigma}}\mathbf{w}
\quad\text{subject to}\quad
\mathbf{1}^{\top}\mathbf{w}=1,
\qquad
\mathbf{w}\geq\mathbf{0}.
\label{eq:longonly}
\end{equation}
The required correction must therefore map the signed spectrum in Eq.~\eqref{eq:pairwise_eigendecomposition} to a positive, regularized spectrum while preserving the information contained in the heterogeneous overlaps.


\section{RIEnet for Incomplete Return Panels}

The architecture follows the analytical GMV workflow of Ref.~\cite{bongiorno2025rienet}. A lag transformation first produces a temporary return panel $\widetilde{\mathbf{R}}$. The resulting sequence then separates into two branches: a shared multilayer perceptron estimates marginal volatilities, while a recurrent spectral module repairs and shrinks the indefinite correlation estimate. The branches recombine into $\boldsymbol{\Sigma}_{\mathrm{NN}}$. The model has approximately 7,400 trainable parameters, independent of $n$ and $\Delta t_{\mathrm{in}}$.

\subsection{Lag Transformation}

Observed returns are mapped elementwise using the same five-parameter lag transformation as in Ref.~\cite{bongiorno2025neural}:
\begin{equation}
\begin{aligned}
\widetilde R_{ti}
&:=
a_t R_{ti}
\frac{\tanh(b_tR_{ti})}{b_tR_{ti}},\\
a_t&:=c_0t^{-c_1},
\qquad
b_t:=c_2-c_3e^{-c_4t},
\end{aligned}
\label{eq:lag_transform}
\end{equation}
$t=1$ denotes the most recent lag, and $c_0,\ldots,c_4>0$ are learned. The coefficient $a_t$ controls the contribution of lag $t$, whereas $b_t$ controls soft clipping. The mask is retained throughout the transformation, so invalid entries do not become observations. The panel $\widetilde{\mathbf{R}}$  is the common input to both subsequent branches.

\subsection{Marginal Volatility}

We apply the marginal-volatility layer of Ref.~\cite{bongiorno2025neural} to the standard deviations in
Eq.~\eqref{eq:marginal_mle}, computed using $\widetilde{\mathbf{R}}$ in place of $\mathbf{R}$. Relative to the
original architecture, the only modification is the inclusion of $T_{ii}^{-1/2}$, which accounts for the asset-specific history length and the resulting heterogeneous sampling uncertainty:
\begin{equation}
\sigma_{i,\mathrm{NN}}
:=
\operatorname{MLP}_{\sigma}
\left(
\begin{bmatrix}
\hat{\sigma}_{i} & T_{ii}^{-1/2}
\end{bmatrix}^{\top}
\right).
\label{eq:volatility_mlp}
\end{equation}
The shared MLP has one hidden layer with $8$ leaky-ReLU units and a softplus output. The resulting scales are divided by the root mean square of the estimated volatilities to have unit mean square. This common normalization does not affect GMV weights.

\subsection{Correlation Estimation}

In the correlation branch, Eqs.~\eqref{eq:marginal_mle}--\eqref{eq:pairwise_eigendecomposition} are evaluated using $\widetilde{\mathbf{R}}$ in place of $\mathbf{R}$, yielding the pairwise cross-moment matrix $\mathbf{C}_{\cap}$, its signed eigenvalues $\lambda_{\cap,k}$, and its eigenvectors $\mathbf{Q}_{\cap}$. Because the entries of $\mathbf{C}_{\cap}$ are estimated from different overlaps, a single panel length cannot characterize the sampling uncertainty of every spectral factor. We transfer the pairwise sample information to factor $k$ through
\begin{equation}
\tau_k
:=
\sum_{i=1}^{n}\sum_{j=1}^{n}
Q_{\cap,ik}^{2}T_{ij}Q_{\cap,jk}^{2}.
\label{eq:factor_effective_length}
\end{equation}
Since the squared eigenvector loadings sum to one, $\tau_k$ is a weighted average of the pairwise overlap counts, where pair $(i,j)$ is weighted by its contribution $Q_{\cap,ik}^{2}Q_{\cap,jk}^{2}$ to factor $k$. It is invariant to eigenvector signs and asset relabeling and reduces to $\Delta t_{\mathrm{in}}$ for a complete panel.

Defining the factor-specific concentration ratio $q_k:=n/\tau_k$, we represent the $k$-th ordered spectral factor by the token
\begin{equation}
\mathbf{x}_k
:=
\left(
\lambda_{\cap,k},\,
\sqrt{n},\,
\sqrt{\tau_k},\,
q_k
\right)^{\top}.
\label{eq:spectral_features}
\end{equation}
The ordered token sequence is processed by a bidirectional GRU following \cite{bongiorno2025rienet}:
\begin{equation}
\overrightarrow{\mathbf{h}}_k
=
\operatorname{GRU}_{\rightarrow}
\left(
\mathbf{x}_k,\overrightarrow{\mathbf{h}}_{k-1}
\right), \quad
\overleftarrow{\mathbf{h}}_k
=
\operatorname{GRU}_{\leftarrow}
\left(
\mathbf{x}_k,\overleftarrow{\mathbf{h}}_{k+1}
\right).
\label{eq:bidirectional_gru}
\end{equation}
Each directional GRU has 32 hidden units.  A shared projection of $\mathbf{h}_k
:=
\left[
\overrightarrow{\mathbf{h}}_k;
\overleftarrow{\mathbf{h}}_k
\right]$ produces a positive inverse eigenvalue:
\begin{equation}
\widetilde{\lambda}_{k,\mathrm{NN}}^{-1}
:=
\operatorname{softplus}
\left(
\boldsymbol{\gamma}^{\top}\mathbf{h}_k+\omega
\right).
\label{eq:spectral_map}
\end{equation}
The inverse eigenvalues are rescaled by a common positive factor so that the corresponding eigenvalues $\lambda_{k,\mathrm{NN}}$ have unit mean. 

The pairwise eigenvectors are rescaled elementwise as
\begin{equation}
Q_{\mathrm{NN},ik}
:=
\frac{Q_{\cap,ik}}
{
\left(
\sum_{\ell=1}^{n}
\lambda_{\ell,\mathrm{NN}}Q_{\cap,i\ell}^{2}
\right)^{1/2}
},
\qquad
\mathbf{C}_{\mathrm{NN}}
:=
\mathbf{Q}_{\mathrm{NN}}
\boldsymbol{\Lambda}_\mathrm{NN}
\mathbf{Q}_{\mathrm{NN}}^{\top}.
\label{eq:cleaned_correlation}
\end{equation}
By construction, $\mathbf{C}_{\mathrm{NN}}$ has unit diagonal and is positive definite. 

\subsection{End-to-End Training}

The marginal and correlation branches are combined to obtain
\begin{equation}
\boldsymbol{\Sigma}_{\mathrm{NN}}
=
\mathbf{D}_{\mathrm{NN}}
\mathbf{C}_{\mathrm{NN}}
\mathbf{D}_{\mathrm{NN}}.
\label{eq:nn_covariance}
\end{equation}
Because $\mathbf{D}_{\mathrm{NN}}$ has positive diagonal entries and $\mathbf{C}_{\mathrm{NN}}$ is positive definite, $\boldsymbol{\Sigma}_{\mathrm{NN}}$ is a valid covariance estimate. During training, we set $\widehat{\boldsymbol{\Sigma}}=\boldsymbol{\Sigma}_{\mathrm{NN}}$ and compute the unconstrained GMV portfolio analytically using Eq.~\eqref{eq:gmv}. The network minimizes its realized five-session variance,
\begin{equation}
\mathcal{L}(\mathbf{w}_{\mathrm{NN}},\boldsymbol{\Sigma}_{\mathrm{out}})
:= n\,
\mathbf{w}_{\mathrm{NN}}
^{\top}
\boldsymbol{\Sigma}_{\mathrm{out}}
\mathbf{w}_{\mathrm{NN}}
.
\label{eq:loss}
\end{equation}
The loss is differentiated through the complete estimation pipeline, from the lag transformation to the reconstructed covariance matrix. 

Although allocation constraints could be incorporated directly into the training problem \cite{lee2025return}, we deliberately train the estimator using the unconstrained analytical solution. Portfolio constraints vary across clients, products, and investment mandates, and embedding a specific constraint set in the loss could require retraining the model whenever the allocation problem changes. We instead learn a covariance estimator independently of the deployment constraints and subsequently apply the common long-only optimizer in Eq.~\eqref{eq:longonly}. This separation evaluates whether the learned risk representation remains useful when the final allocation rule differs from the one used during training.

The architecture is also trained across variable input dimensions. Each training sample contains between 50 and 500 assets ($n$) and between 600 and 1,200 requested history days ( $\Delta t_\mathrm{in}$), whereas the out-of-sample tests contain up to 1,500 assets.  This deliberate dimension shift tests whether the architecture generalizes beyond the cross-sectional sizes observed during training. We train 26 expanding-window models: the first uses 1990--1999 to predict the 2000 out-of-sample year, and the final model uses 1990-2024 for 2025. Training uses Adam with an initial learning rate of $10^{-4}$ and continuous decay $0.99^{b/500}$ after batch $b$, gradient clipping at one, gradient accumulation over two batches of 32 samples, 500 batches per epoch, and 100 epochs.


\section{Experimental Setup}

\subsection{Point-in-Time Investable Universe}

The data cover U.S. common equities and ADRs on the NYSE, Nasdaq, and NYSE American from 1990 through 2025. At each investment date, selection uses only point-in-time information. We retain securities priced between USD 10 and USD 2,000, require at least five million shares outstanding, and remove duplicate company listings by retaining one security per company. Each security must have a complete price history over the most recent 20 sessions. We also exclude securities whose 5- or 20-session log-volatility falls below the cross-sectional lower 1.5-IQR bound, thereby removing near-zero-volatility stocks that could mechanically reduce the GMV objective and trivialize the portfolio problem.  Together, these filters define an investable universe intended to exclude securities with limited execution capacity and to contain transaction costs and market impact, consistent with the use of size, liquidity, and trading-history screens in institutional portfolio construction \cite{msci2026gimi,frazzini2018trading}. The requested universe contains the first 1,500 highest-ranked securities by market capitalization. This construction deliberately retains recently listed and fragmented histories. Across all stock--rebalance observations,  17.15\% have fewer than $\Delta t_{\mathrm{in}}=1,200$ valid returns, 8.45\% have fewer than 600, and 3.48\% have fewer than 252. Unequal return histories are therefore a material feature of the evaluated universe.

Portfolio signals are produced every five trading days using a requested look-back of $\Delta t_\mathrm{in}=1,200$  days. Model inputs consist of the adjusted close-to-close return panel $\mathbf{R}$ and its binary validity mask $\mathbf{M}$, constructed using only information available at the signal time. Returns from the execution day are excluded to avoid look-ahead bias, since the closing price is not yet observed when the portfolio is formed. Every covariance estimator is combined with the same long-only GMV optimizer. Target weights are rounded to increments of $0.1\%$ and renormalized before broker simulation. The out-of-sample period is continuous from January 4, 2000 through December 30, 2025, comprising 6,537 sessions.

\subsection{Compared Estimators}

We first consider estimators designed for incomplete return panels. The nearest-correlation estimator applies an Anderson-accelerated implementation of Higham's projection \cite{anderson1965acceleration,higham2002nearest}. Its bootstrap variant repeats pairwise-complete estimation and nearest-correlation projection over 25 resamples and averages the resulting correlation matrices, thereby adding regularization \cite{bongiorno2024bootstrap}. The factor estimator recovers a low-rank common component from a feasible complete block and uses it to impute missing observations before estimating the covariance matrix \cite{cahan2023factor}. The regression estimator reconstructs the covariance matrix through sequential ridge regressions using the observations available at each step \cite{anderson1957missing,gramacy2008covariance}. We refer to these methods as Anderson, Bootstrap, Factor, and Ridge, respectively.

MLE and Quadratic Inverse Shrinkage (QIS) are instead applied to the complete-row subpanel available at each signal date. Importantly, QIS is derived for a sample covariance computed from a single common panel, with a well-defined observation count and degrees of freedom \cite{ledoitwolf2022qis}. Covariance matrices obtained through pairwise estimation, nearest-correlation projection, or factor imputation do not satisfy this sampling model. Applying RMT after these methods would therefore require a specific statistical derivation.

All covariance estimators use the same long-only GMV optimizer and execution protocol. Equal Weight is included as the allocation-free benchmark. We omit capitalization weighting because it would mechanically concentrate the portfolio in the largest firms and reduce the small-cap exposure central to the experiment.

\subsection{Broker Simulation of Closing-Auction Trading}

The broker simulation starts with USD 1 million and processes the account chronologically. Each session applies splits and dividends, executes the closing rebalance, settles corporate actions, marks positions, and accrues financing. Target holdings are integer shares sized from a strict pre-execution estimate of Net Liquidation Value (NLV) computed using the execution-day opening prices; closing prices are used only for execution. The account ledger, security identifiers, and corporate-action provenance remain continuous through the full 2000-2025 simulation.

Commissions follow the IBKR Pro tiered U.S. equity schedule, including the per-order minimum and notional cap \cite{ibkr2026commissions}; sell orders also incur SEC and FINRA charges. Negative cash accrues financing at the Federal Funds Effective Rate \cite{fred2026dff} plus the broker spread on a 360-day basis. The simulator applies dividends, splits, stock and cash reorganizations, and fractional-share cash-in-lieu.

Let $\Delta S_{ti}$ denote the signed closing-auction order in shares, $\langle V_{ti}\rangle_{10\mathrm{d}}$ the strict-prior ten-day average daily volume, and $P_{ti}$ the raw closing price. Market impact changes the execution price according to
\begin{equation}
P^{\mathrm{exec}}_{ti}
=
P_{ti}
\left[
1+
\operatorname{sign}(\Delta S_{ti})
\kappa_{g(i,t)}
\sqrt{\frac{|\Delta S_{ti}|}{\langle V_{ti}\rangle_{10\mathrm{d}}}}
\right],
\label{eq:impact}
\end{equation}
followed by adverse rounding to the observed price precision. The group $g(i,t)$ is defined by the point-in-time size bucket and listing exchange. The NYSE-like coefficients $\kappa$ are 0.45\%, 0.74\%, and 1.82\% for large, small, and micro stocks; Nasdaq adds 0.57, 0.51, and 2.66 percentage points, respectively \cite{goyal2026impact}. These coefficients enter only the ex post execution simulation and do not affect covariance estimation, optimization, or target-weight formation.



\section{Results}

\begin{table}[t]
  \centering
  \caption{Out-of-sample performance with physical execution, January 2000--December 2025. Vol. 5d is annualized from non-overlapping five-session returns; $\beta_{\mathrm{RUI}}$ is estimated at the same horizon against the Russell 1000 price index; MCS $p$-values use squared five-session-return losses. All figures are net of execution costs. Methods are ordered by increasing five-session volatility.}
  \label{tab:physical-execution-performance}
  \small
  \setlength{\tabcolsep}{5.0pt}
  \begin{tabular}{lrrrrrr}
    \toprule
    Method
      & Vol. 5d (\%)
      & CAGR (\%)
      & Sharpe
      & Max DD (\%)
      & $\beta_{\mathrm{RUI}}$
      & MCS $p$ \\
    \midrule
    RIEnet    & \textbf{11.17} & \textbf{9.30} & \textbf{0.814} & \textbf{-41.3} & \textbf{0.500} & \textbf{1.000} \\
    Factor    & 14.08 & 7.33 & 0.580 & -51.5 & 0.592 & $<0.001$ \\
    Bootstrap & 14.46 & 7.22 & 0.563 & -49.4 & 0.595 & $<0.001$ \\
    Anderson  & 14.74 & 7.55 & 0.578 & -50.8 & 0.603 & $<0.001$ \\
    QIS       & 15.54 & 8.07 & 0.560 & -50.3 & 0.799 & $<0.001$ \\
    MLE       & 16.25 & 2.75 & 0.248 & -60.3 & 0.683 & $<0.001$ \\
    Ridge     & 18.07 & 3.27 & 0.272 & -70.0 & 0.671 & $<0.001$ \\
    \midrule
    Equal Weight & 20.90 & 7.53 & 0.440 & -58.5 & 1.125 & $<0.001$ \\
    \bottomrule
  \end{tabular}
  \Description{Comparison of annualized five-day volatility, net CAGR, Sharpe ratio, maximum drawdown, five-session Russell 1000 beta, and loss-based Model Confidence Set p-values for seven covariance estimators and the Equal Weight benchmark.}
\end{table}

Table~\ref{tab:physical-execution-performance} reports out-of-sample performance after broker-simulated execution and all modeled trading costs. Five-session realized volatility is the primary endpoint because it is aligned with the horizon of the training objective and the portfolio rebalancing frequency. RIEnet attains the lowest annualized volatility, $11.17\%$, compared with $14.08\%$ for Factor, the second-best covariance estimator. This corresponds to a $20.7\%$ reduction. We assess the statistical significance of the realized-risk differences using the Model Confidence Set (MCS) of Hansen et al.~\cite{hansen2011mcs} with a stationary block bootstrap \cite{politis1994stationary}. Using $10{,}000$ bootstrap replications, RIEnet is the only method retained at the $0.1\%$ significance level for the baseline expected block length of $11$ and for all sensitivity values, $5$, $10$, $20$, and $40$ blocks.

\begin{figure}[t]
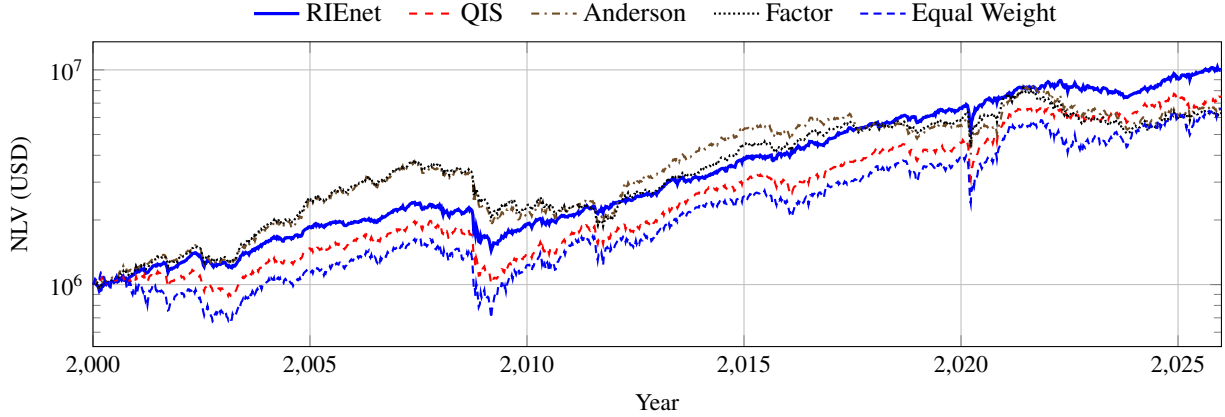

  \centering

  \caption{NLV under broker-simulated closing-auction execution on a logarithmic scale. QIS represents complete-row RMT shrinkage, Anderson nearest-correlation projection, Factor incomplete-panel imputation, and Equal Weight the non-estimation benchmark. Curves are sampled at the five-session rebalancing frequency; reported statistics use daily data.}
  \Description{Log-scale NLV from 2000 through 2025 for RIEnet, QIS, Anderson, Factor, and Equal Weight. RIEnet ends at the highest NLV.}
  \label{fig:equity-curve}
\end{figure}

The secondary performance measures show that the reduction in realized risk is not obtained by sacrificing portfolio growth. RIEnet achieves the highest CAGR, $9.30\%$, and the highest Sharpe ratio, $0.814$, compared with $8.07\%$ and $0.560$ for QIS and $7.33\%$ and $0.580$ for Factor. Relative to the next-highest Sharpe ratio, the improvement is approximately $40\%$. RIEnet also records the least severe maximum drawdown among the covariance estimators, at $-41.3\%$, compared with values between $-49.4\%$ and $-70.0\%$ for the alternatives. Figure~\ref{fig:equity-curve} shows that these aggregate differences are not generated by a few favorable episodes. RIEnet maintains a stable performance advantage across the full 26-year experiment and through the principal market drawdowns represented in the sample. The separation becomes particularly pronounced in the final years and does not exhibit the late-sample deterioration that would indicate temporal overfitting. This evidence is especially relevant for a machine-learning estimator, since instability under changing market conditions would be expected to become most visible in the most recent out-of-sample years \cite{fischer2018deep}.

RIEnet also has the lowest sensitivity to the Russell 1000 among the evaluated covariance estimators. Its five-session beta is $0.500$, compared with $0.592$ for Factor, $0.799$ for QIS, and $1.125$ for Equal Weight. The improvement in realized risk and compounded performance is therefore not obtained by mechanically reproducing the exposure of a broad capitalization-weighted equity index.

\begin{table}[t]
  \centering
  \caption{Execution diagnostics. The $n_{\mathrm{eff}}$ column reports the median effective number of assets across rebalancing dates \cite{woerheide1992index}. Turnover is annualized gross traded notional divided by pre-trade NLV; impact and fees are basis points per dollar traded; CAGR drags are annual percentage points. Covariance estimators are ordered by increasing total drag; Equal Weight is reported separately. Boldface identifies the best optimized-portfolio value in each execution-cost column.}
  \label{tab:execution-cost-diagnostics}
  \small
  \setlength{\tabcolsep}{1.45pt}
  \begin{tabular}{@{}lrrrrrr@{}}
    \toprule
    Method
      & $n_{\mathrm{eff}}$
      & Turnover
      & Impact
      & Fees
      & Impact drag
      & Total drag \\
    \midrule
    QIS       & 754.1 & 24.60 & \textbf{0.87} & 2.48 & \textbf{0.27} & \textbf{1.17} \\
    RIEnet    & 57.6  & 31.53 & 2.57 & \textbf{0.92} & 1.09 & 1.49 \\
    Factor    & 22.5  & \textbf{19.24} & 12.33 & 1.41 & 2.74 & 3.06 \\
    Ridge     & 14.6  & 32.58 & 7.85 & 1.49 & 2.93 & 3.46 \\
    MLE       & 34.7  & 59.50 & 3.72 & 1.33 & 2.79 & 3.68 \\
    Anderson  & 19.1  & 26.07 & 11.55 & 1.29 & 3.50 & 3.90 \\
    Bootstrap & 19.7  & 28.76 & 10.77 & 1.27 & 3.60 & 4.03 \\
    \midrule
    Equal Weight & 1496.8 & 3.93 & 0.32 & 17.79 & 0.01 & 1.01 \\
    \bottomrule
  \end{tabular}
  \Description{Median effective number of assets, annualized turnover, market-impact and fee rates, and their estimated annual CAGR drag for seven covariance estimators and the Equal Weight benchmark.}
\end{table}

The portfolio-concentration and execution diagnostics clarify how these statistical gains translate into implementation costs. Table~\ref{tab:execution-cost-diagnostics} reports the median effective portfolio size computed from target weights, together with annualized turnover, market impact, explicit fees, and their estimated effect on compounded performance. RIEnet has a median effective size of $57.6$ positions, making it substantially more diffuse than MLE, Bootstrap, Anderson, Factor, and Ridge, but considerably more concentrated than QIS, whose median effective size is $754.1$. RIEnet turns over $31.53$ times per year and incurs $2.57$ basis points of market impact and $0.92$ basis points of explicit fees per dollar traded, corresponding to an estimated total CAGR drag of $1.49$ percentage points.

QIS has the lowest total drag among the optimized portfolios, at $1.17$ percentage points, consistent with its broad allocation and smaller individual orders. RIEnet is therefore not the least costly estimator to implement, but its total drag remains well below those of MLE, Bootstrap, Anderson, Factor, and Ridge, which range from $3.06$ to $4.03$ percentage points. The results show that low turnover does not necessarily imply low implementation costs, since concentrated trades can generate substantial market impact. Execution costs therefore depend jointly on portfolio breadth, turnover, and order concentration. RIEnet preserves its realized-risk and risk-adjusted-performance advantage after broker-simulated execution while incurring substantially lower implementation costs than the more concentrated incomplete-panel estimators.

\section{Discussion and Conclusions}

Incomplete return panels arise naturally in equity universes that include recent listings, lower-capitalization securities, and histories of unequal length. Pairwise-complete estimation preserves this information, but introduces heterogeneous sampling uncertainty and may produce indefinite correlation matrices. The proposed estimator addresses both issues jointly by conditioning a rotation-invariant spectral map on factor-aligned overlap information and transforming the signed spectrum into a valid, regularized covariance estimate.

The empirical results indicate that this treatment is economically relevant. RIEnet achieves the lowest realized risk and the highest risk-adjusted performance among the evaluated covariance estimators, remains the only method retained by the Model Confidence Set, and preserves its advantage after the modeled execution costs. These gains are not obtained by collapsing toward a broadly diversified allocation: the resulting portfolios remain substantially more selective than QIS and exhibit the lowest Russell 1000 beta among the considered covariance estimators.

The empirical analysis evaluates the fixed RIEnet specification within
the full portfolio-construction and execution pipeline. Extended
component-level attribution is most cleanly conducted first on
controlled synthetic benchmarks, where the population covariance and
missingness mechanism are known and architectural effects can be
isolated without repeatedly reusing the historical backtest
\cite{gu2020empirical,white2000reality}. We leave this broader analysis
to future work.



\end{document}